\documentclass[mathleft
]{an}
\usepackage{graphicx}
\usepackage{times}
\usepackage{tabularx}
\usepackage{amsmath}
\usepackage[varg]{txfonts}
\usepackage{txfonts}
\usepackage{wasysym}
\usepackage[english]{babel}
\begin{document}

\Pagespan{789}{}
\Yearpublication{2006}%
\Yearsubmission{2005}%
\Month{11}%
\Volume{999}%
\Issue{88}%

\title{TASTE IV. Refining ephemeris and orbital parameters \\for HAT-P-20b and WASP-1b}

\author{V.~Granata\inst{1,2}\fnmsep\thanks{Corresponding author:
  \email{valentina.granata@unipd.it}\newline}
	  \and V.~Nascimbeni\inst{1,2}
	  \and G.~Piotto\inst{1,2}
	  \and L.~R.~Bedin\inst{2}
	  \and L.~Borsato\inst{1,2}
	  \and A.~Cunial\inst{1,2}
	  \and M.~Damasso\inst{1,3}
	  \and L.~Malavolta\inst{1,2}
          }
\titlerunning{TASTE IV. Refining ephemeris and orbital parameters for HAT-P-20b and WASP-1b}
\authorrunning{V. Granata \textit{et~al.}}
\institute{
Dipartimento di Fisica e Astronomia, Universit\`a degli Studi di Padova,
Vicolo dell'Osservatorio 3, 35122 Padova, Italy
\and 
INAF -- Osservatorio Astronomico di Padova, vicolo dell'Osservatorio 5, 35122 Padova, Italy
\and 
Astronomical Observatory of the Autonomous Region of the Aosta Valley, 
             Loc.\ Lignan 39, 11020 Nus (AO), Italy
}

\received{3 March 2014}
\accepted{13 May 2014}
\publonline{later}


  \abstract{We present four new light curves of transiting exoplanets WASP-1b and HAT-P-20b, 
observed within the TASTE (\emph{The Asiago Search for Transit timing variations of Exoplanets}) project.
 We re-analyzed light curves from the literature in a homogeneous way, calculating a refined ephemeris 
and orbital-physical parameters for both objects. 
 WASP-1b does not show any significant Transit Timing Variation signal at the 120\,s-level. As for HAT-P-20b, we detected a 
deviation from our re-estimated linear ephemeris that could be ascribed to the presence of a perturber or, 
more probably, to a previously unnoticed high level of stellar activity. The rotational period of 
HAT-P-20~A we obtained from archival data ($P_\mathrm{rot}\simeq 14.5$ days), combined with its optical 
variability and strong emission of CaII H\&K lines, is consistent with a young stellar age ($< 1$ Gyr) 
and support the hypothesis that stellar activity may be responsible of the measured deviations of the 
transit times.
} 

\maketitle

\section{Introduction}

A promising method to detect exoplanets is the transit technique paired with the Transit Timing Variation (TTV) analysis.
It is well known that a planet transiting on its star in a Keplerian orbit produces a strictly periodic signal only if it is unperturbed by other bodies. If a third companion is present, transits are no longer exactly periodic (Holman \& Murray, 2005). In a three-body system, the mass of the perturber can be inferred by measuring the timing deviations of the central instants of the transits from the expected linear ephemeris (Agol et~al.~2005).

The TASTE project (\emph{The Asiago Search for Transit timing variations of Exoplanets}, Nascimbeni et~al. 2011a, Paper~I hereafter) is focused on the TTV method. The aim of the program is to collect a database of high-precision light curves to be searched for timing variations exploiting the Asiago Observatory\footnote{Based on observations collected at Copernico telescope (Asiago, Italy) of the INAF - Osservatorio Astronomico di Padova.} instruments (Paper~I; Nascimbeni et~at. 2011b, Paper~II hereafter) and, more recently, other medium-class facilities around the world.

In this work, we present a new transit light curve of WASP-1b and three additional light curves of HAT-P-20b, also re-analyzing data published from different authors in order to provide a new reference ephemeris for both planets.

WASP-1b (Collier Cameron et~al. 2007) is a hot Jupiter transiting an F7V host star ($V = 11.79$) with a period of $P = 2.51995 \pm 0.000001$ days, a radius of $R_\mathrm{p} \simeq 1.3$ $R_\mathrm{jup}$, and a derived mass $M_p\simeq0.86$ $M_\mathrm{jup}$.

HAT-P-20b (Bakos et~al. 2011) is a massive and extremely dense transiting exoplanet ($M_\mathrm{p}\simeq 7.2 M_\mathrm{jup}$, $R_\mathrm{p} \simeq 0.87$ $R_\mathrm{jup}$) orbiting around a $V = 11.339$ K3V star with a period $P = 2.875317 \pm 0.000004$ days. The parent star HAT-P-20~A is the primary component of a visual binary system (Bakos et~al. 2011). The companion HAT-P-20~B is a common-proper-motion red dwarf at $\sim 7^{\prime\prime}$ from HAT-P-20~A and, according to the same authors, is fainter by 1.36 mag in a unspecified $R$ band.
HAT-P-20~A has been recently recognized (Ballerini et~al. 2012, Oshagh et~al. 2013) as a possibly spotted star and, as we discuss in this work, its activity could be linked to the anomalous scatter we found in the $O-C$ diagram.

\begin{figure*}
\centering
\includegraphics[width=0.75\textwidth, clip]{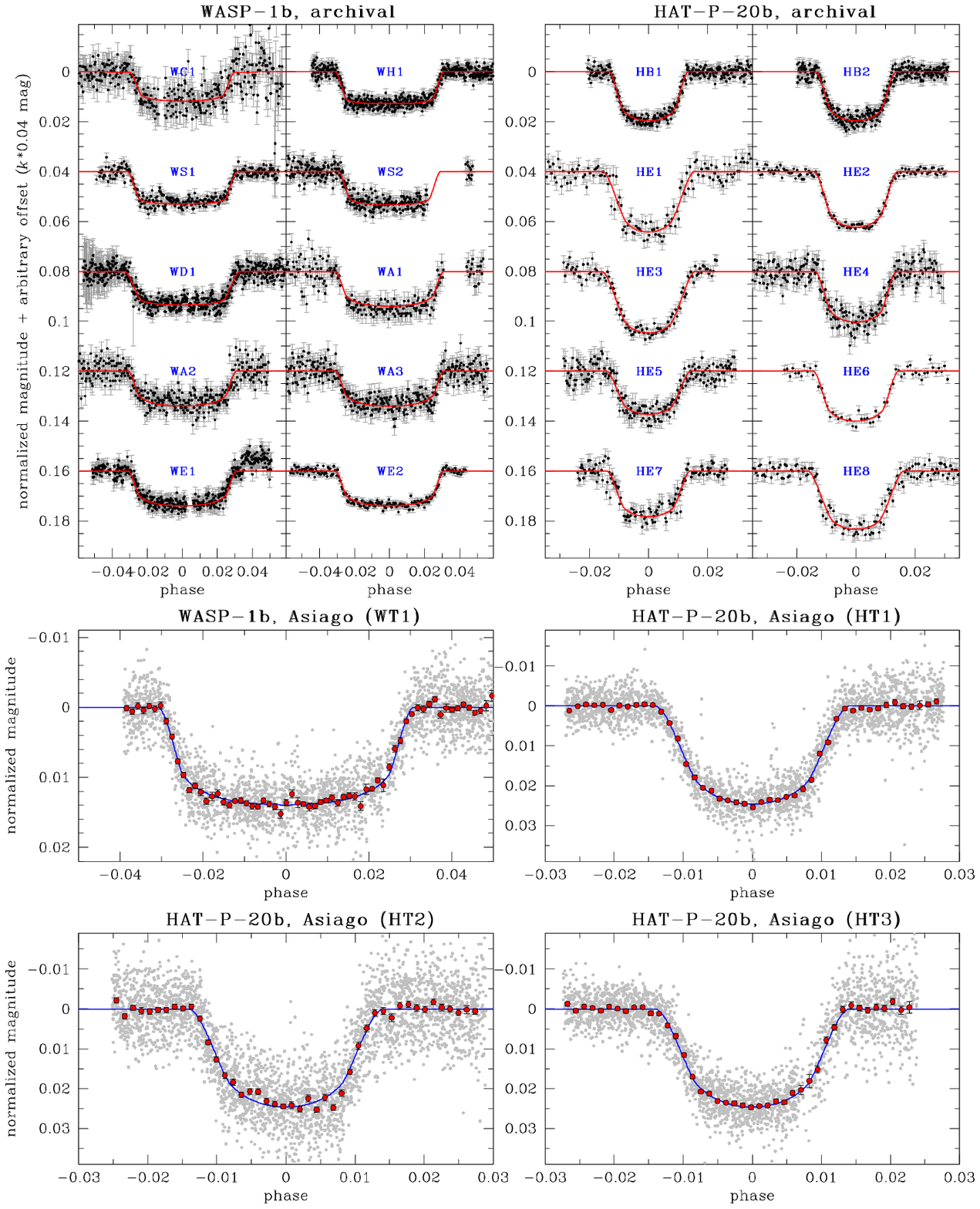}
\caption{New TASTE transits of WASP-1b (\texttt{WT1}) and HAT-P-20b (\texttt{HT1}, \texttt{HT2}, \texttt{HT3}) acquired at the Asiago 1.82m Copernico telescope. Gray dots represent the unbinned points with the original time sampling (respectively: 7\,s for \texttt{WT1}, 5\,s for \texttt{HT1} and \texttt{HT3}, and 4\,s for \texttt{HT2}), while red dots are binned on 2-min intervals. Error bars on the latter are computed as the standard error within each bin of the residuals from the model. }
\label{lc}
\end{figure*}

\section{Observations} \label{sec2}

The new planetary transits were observed with the Asiago 1.82m ``Copernico'' telescope. 
A complete description of the TASTE instrumental setup, observing strategy, data reduction/analysis process, and the monitored sample can be found in Paper~I.

We observed WASP-1b on the night of 2010 November 29, while HAT-P-20b was observed on 2011 February 5 (a few clouds appeared after 200 minutes from the start, after the last contact), November 28 (the sky was clear except for the off-transit part, between $150$-$200$ minutes from the start), and December 22 (some clouds affected the series during the first 20 minutes before the ingress and during the second half of the transit). In the following sections we will refer to these light curves with the ID codes \texttt{WT1}, \texttt{HT1}, \texttt{HT2}, \texttt{HT3}, respectively.

For all four photometric series, we employed a Cousins $R_\textrm{C}$ filter both to maximize the number of detected photons and to minimize the atmospheric extinction and limb darkening effects. Stellar images were defocused to a FWHM of $\sim5$-$8$ pixels (the spanning was due to focus changes) to prevent saturation and minimize pixel-to-pixel systematic errors.

The summary of our WASP-1b and HAT-P-20b observations is given in Table~\ref{wasp1summary} and Table~\ref{hat20summary}.

\section{Data reduction and analysis} \label{reduction}

\subsection{Data reduction}
The entire data reduction procedure and the routines handled decorrelation of systematic errors as a function of airmass, centroid positions, FWHM, and other quantities of interest adopted here are described in Paper~I. The TASTE light curves for both targets are shown in Fig.~\ref{lc}.

\begin{figure}
\centering
\includegraphics[width=0.95\columnwidth]{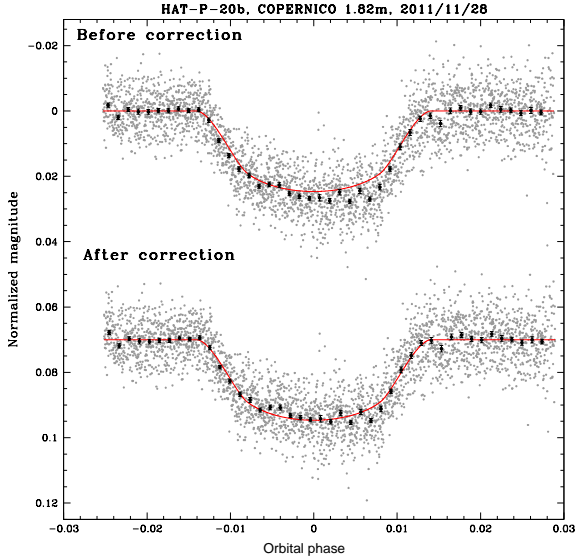}
\caption{Light curve of HAT-P-20b before and after the correction due to stellar companion pollution for the transit event on 2011 November 28. The bottom light curve is the same as \texttt{HT2} in Fig.~\ref{lc}.}
\label{befaft}
\end{figure}

Upon visual inspection, we noticed a systematic effect in the \texttt{HT2} light curve only (Fig.~\ref{befaft}, \emph{upper panel}). We ran some tests to investigate the possible source of this effect. It turned out that the light curve of the primary companion, HAT-P-20~A, is polluted by variable contaminating flux coming from the close companion HAT-P-20~B.
We re-evaluated the light curve with an oversized aperture (11$^{\prime\prime}$ corresponding to a radius of 23 pixels) centered on HAT-P-20~A, to include all the light from both stellar sources. Then we estimated the off-transit flux ratio between A and B components through an undersized 2.5$^{\prime\prime}$ (5 pixels) aperture centered on primary component obtaining $f_B/f_A = 0.1524 \pm 0.0011$ (corresponding to a dilution factor of about 87\%) and applied this correction to \texttt{HT2} data only, restoring the original transit depth.
The relative light curve after the correction for \texttt{HT2} is shown in the lower panel of Fig.~\ref{befaft}.

\subsection{Data analysis}
We re-analyzed all the archival data published in the refereed literature, and also some light curves by amateur astronomers stored in the ETD\footnote{http://var2.astro.cz/ETD/index.php} and AXA\footnote{http://nsted.ipac.caltech.edu/NStED/docs/datasethelp/AXA.html} archives. Only complete light curves with a maximum error in depth $< 5$\% have been selected. Among those, we discarded the Szab\'o et~al. (2010) observation of WASP-1b because the measurements are not reliable due to large error bars which hamper any TTV analysis, according to what the author themselves declared in their paper. The published $T_0$ is in disagreement by $> 6\,\sigma$ with other professional measurements carried out in the same season. 

Employing \texttt{JKTEBOP} code\footnote{http://www.astro.keele.ac.uk/$\sim$jkt/codes/jktebop.html} (version 25, Southworth, Maxted \& Smalley, 2004), we fitted a transit model to our light curves fixing the quadratic limb darkening coefficient $u_2$ to its theoretical value, while the linear parameter $u_1$ was guessed, as proposed by Southworth (2010). 
The resulting best-fit models for both targets, normalized to their off-transit magnitude, are represented with blue curves in Fig.~\ref{lc}.
This choice turned out to be the best approach, providing the minimum reduced $\chi^2$ value and estimating realistic uncertainties on fitted parameters. We took theoretical limb darkening coefficients from the \texttt{JKTLD} code\footnote{http://www.astro.keele.ac.uk/$\sim$jkt/codes/jktld.html}, interpolating the Claret (2000) tables. The orbital inclination $i$, fractional ratio $R_\mathrm{p}/R_\star$, and sum of fractional radii $(R_\mathrm{p} + R_\star)/a$ were then fitted on photometric series along with $u_1$ and the central transit time $T_0$. As the linear limb darkening coefficient depends on wavelength, no average value for this parameter was calculated for this heterogeneous data set.
Table~\ref{wasp1summary} and Table~\ref{hat20summary} summarize relevant information about the whole data set, in particular the newly-estimated best-fit values of $T_0$ (BJD$_\mathrm{TDB}$, the Barycentric Julian Date in Barycentric Dynamical Time, Eastman, Siverd, \& Gaudi 2010), for each transit of WASP-1b and HAT-P-20b.
The individual fitted values are summarized on Table~\ref{param}. 
In all Tables, a unique three-character ID code is assigned to each light curve. 

In some cases, archival light curves were secured using filters other than $R$, or without filter (referred as ``clear'' in Tables~\ref{wasp1summary} and \ref{hat20summary}): in the former case we interpolated the limb darkening coefficients from the Claret (2000, 2004) tables as above; in the latter, we computed the average value between $V$ and $R$ bands given by Claret (2000), assuming that the quantum efficiency of a typical unfiltered CCD peaks somewhere between these bands (Nascimbeni et~al. 2013, Paper~III hereafter).

Residual trends in a few archival light curves were removed by subtracting a low-order polynomial fitted over the off-transit part. Where photometric uncertainties are not given in the raw data, the off-transit RMS was adopted as constant photometric error. Before the fitting procedure, the time coordinate of all light curves has been uniformly converted to the BJD$_\mathrm{TDB}$ standard time, just as for the Asiago data.

We estimated the errors through a residual-shift method. It takes into account the presence of red noise in the light curve by propagating any systematic error along the series while preserving its structure (Southworth 2008, Paper~III, and references therein for more details).

\begin{itemize}
\item \emph{WASP-1b.} The photometric scatter for \texttt{WT1}, measured as the 68.27$^{\mathrm{th}}$ percentile of the residuals from the best-fit model, is $\sigma_\mathrm{u} = 2.7$ mmag (unbinned points). 
By definition, the net cadence (or the time sampling) $\tau$ is the ratio between the total time of the photometric series and the number of acquired frames, where the total time of the series is the sum of the exposure and the reading time. For \texttt{WT1}, $\tau \sim 7$\,s.
These quantities enable us to calculate the corresponding 2-min binned RMS $\sigma_{120} = \sigma_\mathrm{u}\sqrt{\tau/120~\mathrm{s}}$. The $\sigma_{120}$ values for all the analyzed light curves are shown in Table~\ref{wasp1summary}.
All the estimated values of $R_\mathrm{p}/R_\star$ for WASP-1b are consistent within their error bars with the corresponding estimate published in the original works, with the exception of \texttt{WC1} which has the lowest S/N among the data set, making the fit procedure unstable. For this reason, we did not employ this transit in the subsequent evaluation of the weighted means of $i$, $(R_\mathrm{p} + R_\star)/a$ and $R_\mathrm{p}/R_\star$ (last row of upper sub-table of Table~\ref{param}) as final best-fit values.
We note that the errors on our final weighted means are significantly smaller than any other estimate published so far.
\item \emph{HAT-P-20b.} The measured scatter $\sigma_\mathrm{u}$ is respectively $3.6$ mmag with a $\tau \sim 5$\,s (\texttt{HT1}), $6.3$ mmag  with a $\tau \sim 4$\,s (\texttt{HT2}), and $5.3$ mmag  with a $\tau \sim 5$\,s (\texttt{HT3}), while the corresponding $\sigma_{120}$ values are shown in Table~\ref{hat20summary}. For this object, we calculated the weighted means of $i$ and $(R_\mathrm{p} + R_\star)/a$ only (last row of lower sub-table of Table~\ref{param}) but not for the ratio $R_\mathrm{p}/R_\star$, whose estimate appears to disagree with each other. The possible reasons for that are discussed in Section~\ref{discussion}.
\end{itemize}
It is worth mentioning that TASTE $\sigma_{120}$ values for both targets are the smallest amongst the relative two samples: this is probably due to the larger collecting area of the Copernico telescope and to our defocusing stategy.
\begin{table*}
\caption{Summary of the light curves of WASP-1b analyzed in this work. }
\label{wasp1summary}
\centering
\scalebox{0.74}{
\setlength{\extrarowheight}{8pt}
\begin{tabular}{rcccccllccl}
\hline
\raisebox{1ex}[0pt][0pt]{$N$}& \raisebox{1ex}[0pt][0pt]{Date} &\raisebox{1ex}[0pt][0pt]{$T_0$ (best-fit) $\pm \sigma$ } & \raisebox{1ex}[0pt][0pt]{$T_0$ (med) } & \raisebox{1ex}[0pt][0pt]{$O-C$ } & \raisebox{1ex}[0pt][0pt]{$\frac{O-C}{\sigma}$}& \raisebox{1ex}[0pt][0pt]{Telescope} &\raisebox{1ex}[0pt][0pt]{Filter}& \raisebox{1ex}[0pt][0pt]{$\sigma_{120}$}& \raisebox{1ex}[0pt][0pt]{ID} & \raisebox{1ex}[0pt][0pt]{Reference}\\
&&\raisebox{1ex}[0pt][0pt]{(BJD$_{\mathrm{TDB}}$)}&\raisebox{1ex}[0pt][0pt]{(BJD$_{\mathrm{TDB}}$)}& \raisebox{1ex}[0pt][0pt]{(days)}&&&&\raisebox{1ex}[0pt][0pt]{(mmag)}&\\
\hline
$-$642 &2006/10/02& $ 2454010.78632 \pm 0.00175 $ & $ 2454010.78621 ^{\mathstrut + 0.00186} _{\mathstrut - 0.00164} $ &  $ -0.00571 $ & $ -3.59 $ & KO-0.35m& $R$&3.4& \texttt{WC1} & Cameron et~al. (2007)\\
$-$605 &2006/09/27& $ 2454005.75274 \pm 0.00038 $ & $ 2454005.75269 ^{\mathstrut + 0.00051} _{\mathstrut - 0.00026} $ &  $ -0.00036 $ & $ -0.95 $ & FLWO-1.2m &Sloan $z$&1.0& \texttt{WH1} & Charbonneau et~al. (2007)\\
$-$602 &2006/10/04& $ 2454013.31335 \pm 0.00045 $ & $ 2454013.31332 ^{\mathstrut + 0.00050} _{\mathstrut - 0.00040} $ &  $\phantom{-}  0.00041 $ & $ \phantom{-}0.91 $ & Wise-1.0m& $I$&1.5& \texttt{WS1} & Shporer et~al. (2007)\\
$-$600 &2006/10/09& $ 2454018.34984 \pm 0.00282 $ & $ 2454018.34975 ^{\mathstrut + 0.00303} _{\mathstrut - 0.00261} $ &  $ -0.00299 $ & $ -1.06 $ & Wise-1.0m& $I$&1.9& \texttt{WS2} & Shporer et~al. (2007)\\
$-$596 &2006/10/19& $ 2454028.43403 \pm 0.00179 $ & $ 2454028.43448 ^{\mathstrut + 0.00227} _{\mathstrut - 0.00133} $ &  $\phantom{-}  0.00143 $ & $ \phantom{-}0.80 $ & Finland-0.30m& $R$&2.9& \texttt{WA1} &AXA obs.~Hentunen\\
$-$453 &2007/10/15& $ 2454388.78506 \pm 0.00089 $ & $ 2454388.78508 ^{\mathstrut + 0.00096} _{\mathstrut - 0.00081} $ &  $\phantom{-}  0.00035 $ & $ \phantom{-}0.40 $ & Arizona-0.36m& $R$&2.6& \texttt{WA2} &AXA obs.~Gary	\\ 
$-$315 &2008/09/26& $ 2454736.53888 \pm 0.00182 $ & $ 2454736.53887 ^{\mathstrut + 0.00186} _{\mathstrut - 0.00178} $ &  $\phantom{-}  0.00179 $ & $ \phantom{-}0.98 $ & T1T-1.20m& $R$&1.5& \texttt{WE1} &ETD obs.~Olhert\\
$-$305 &2008/10/22& $ 2454761.73594 \pm 0.00066 $ & $ 2454761.73598 ^{\mathstrut + 0.00061} _{\mathstrut - 0.00071} $ &  $ -0.00060 $ & $ -0.91 $ & KPNO-2.1m& $J$&1.5& \texttt{WD1} & Sada et~al. (2012)\\
$-$305 &2008/10/22& $ 2454761.73804 \pm 0.00102 $ & $ 2454761.73793 ^{\mathstrut + 0.00100} _{\mathstrut - 0.00103} $ &  $\phantom{-}  0.00150 $ & $ \phantom{-}1.48 $ & Arizona-0.28m& clear	&2.5& \texttt{WA3} &AXA obs.~Gary	\\
$-$38  &2010/08/25& $ 2455434.56159 \pm 0.00024 $ & $ 2455434.56162 ^{\mathstrut + 0.00017} _{\mathstrut - 0.00030} $ &  $ -0.00021 $ & $ -0.88 $ & Calar Alto-1.23m& $R$&1.0& \texttt{WE2} &ETD obs.~Hormuth\\
    0  &2010/11/29& $ 2455530.31981 \pm 0.00019 $ & $ 2455530.31983 ^{\mathstrut + 0.00016} _{\mathstrut - 0.00022} $ &  $\phantom{-}0.00011 $ & $ \phantom{-}0.58 $ & Asiago-1.82m& $R_\textrm{C}$&0.7& \texttt{WT1} &This work\\
\hline
\end{tabular}
}
\note{The columns give: the transit epoch $N$ assuming $T=T_0+NP$ and the new ephemeris Eq.\,(\ref{t0_WT1}), the ``evening date'' of the observation, the best-fit value for the central instant $T_0$ of the transit and the associated $\sigma$, the median value of the distribution of $T_0$ from the Residual Permutation (RP) algorithm and the associated errors $\sigma_+$ and $\sigma_-$, the $O-C$ according to Eq.\,(\ref{t0_WT1}) in days, the $O-C$ in units of $\sigma$, the telescope and filter employed, the normalized photometric scatter $\sigma_{120}$, the ID code of the light curve, and the reference paper or database.}
\end{table*}
\begin{table*}
\caption{Summary of the light curves of HAT-P-20b analyzed in this work. }
\label{hat20summary}
\centering
\scalebox{0.75}{
\setlength{\extrarowheight}{8pt}
\begin{tabular}{rcccccllccl}
\hline
\raisebox{1ex}[0pt][0pt]{$N$}& \raisebox{1ex}[0pt][0pt]{Date} &\raisebox{1ex}[0pt][0pt]{$T_0$ (best-fit) $\pm \sigma$ } & \raisebox{1ex}[0pt][0pt]{$T_0$ (med) } & \raisebox{1ex}[0pt][0pt]{$O-C$ } & \raisebox{1ex}[0pt][0pt]{$\frac{O-C}{\sigma}$}& \raisebox{1ex}[0pt][0pt]{Telescope} &\raisebox{1ex}[0pt][0pt]{Filter}& \raisebox{1ex}[0pt][0pt]{$\sigma_{120}$}& \raisebox{1ex}[0pt][0pt]{ID} & \raisebox{1ex}[0pt][0pt]{Reference}\\
&&\raisebox{1ex}[0pt][0pt]{(BJD$_{\mathrm{TDB}}$)}&\raisebox{1ex}[0pt][0pt]{(BJD$_{\mathrm{TDB}}$)}& \raisebox{1ex}[0pt][0pt]{(days)}&&&&\raisebox{1ex}[0pt][0pt]{(mmag)}&\\
\hline
$-$242 &2009/03/11& $ 2454902.65764 \pm 0.00019 $ & $ 2454902.65761 ^{\mathstrut + 0.00022} _{\mathstrut - 0.00016} $ & $ -0.00008 $ & $ -0.40 $ & FLWO-1.2m& Sloan $i$ &1.3& \texttt{HB1} & Bakos et~al. (2011)\\
$-$164 &2009/10/21& $ 2455126.93245 \pm 0.00019 $ & $ 2455126.93245 ^{\mathstrut + 0.00018} _{\mathstrut - 0.00020} $ & $ -0.00012 $ & $ -0.63 $ & FLWO-1.2m& Sloan $i$ &1.1	& \texttt{HB2} & Bakos et~al. (2011)\\
  $-$8 &2011/01/13& $ 2455575.47974 \pm 0.00076 $ & $ 2455575.47973 ^{\mathstrut + 0.00066} _{\mathstrut - 0.00086} $ & $ -0.00253 $ & $ -3.33 $ & SC-0.30m& $R$ &3.6&\texttt{HE1} &ETD obs.~Naves\\
  $-$8 &2011/01/13& $ 2455575.48039 \pm 0.00037 $ & $ 2455575.48038 ^{\mathstrut + 0.00039} _{\mathstrut - 0.00036} $ & $ -0.00188 $ & $ -5.09 $ & SC-0.35m& clear &1.3& \texttt{HE2} &ETD obs.~Garcia\\
     0 &2011/02/05& $ 2455598.48522 \pm 0.00016 $ & $ 2455598.48524 ^{\mathstrut + 0.00012} _{\mathstrut - 0.00020} $ & $  \phantom{-}0.00040 $ & $ \phantom{-}2.49 $ & Asiago-1.82m& $R_\textrm{C}$ &0.8	& \texttt{HT1} &This work\\
     1 &2011/02/08& $ 2455601.35956 \pm 0.00036 $ & $ 2455601.35948 ^{\mathstrut + 0.00043} _{\mathstrut - 0.00029} $ & $ -0.00058 $ & $ -1.61 $ & Newton-0.25m& clear &2.3	& \texttt{HE3} &ETD obs.~P$\check{\textrm{r}}$ib\'ik\\
   103 &2011/11/28& $ 2455894.64292 \pm 0.00046 $ & $ 2455894.64292 ^{\mathstrut + 0.00047} _{\mathstrut - 0.00044} $ & $  \phantom{-}0.00028 $ & $ \phantom{-}0.61 $ & Newton-0.20m& clear &2.8	& \texttt{HE4} &ETD obs.~Trnka\\
   103 &2011/11/28& $ 2455894.64393 \pm 0.00043 $ & $ 2455894.64400 ^{\mathstrut + 0.00038} _{\mathstrut - 0.00049} $ & $  \phantom{-}0.00129 $ & $ \phantom{-}3.00 $ & Asiago-1.82m& $R_\textrm{C}$ &1.2	& \texttt{HT2} &This work\\
   104 &2011/12/01& $ 2455897.51844 \pm 0.00074 $ & $ 2455897.51856 ^{\mathstrut + 0.00062} _{\mathstrut - 0.00087} $ & $  \phantom{-}0.00048 $ & $ \phantom{-}0.65 $ & Refractor-0.16m& clear &2.6	& \texttt{HE5} &ETD obs.~Ayomamitis\\
   111 &2011/12/22& $ 2455917.64445 \pm 0.00028 $ & $ 2455917.64444 ^{\mathstrut + 0.00043} _{\mathstrut - 0.00013} $ & $ -0.00074 $ & $ -2.64 $ & Asiago-1.82m& $R_\textrm{C}$ &0.9	& \texttt{HT3} &This work\\
   129 &2012/02/11& $ 2455969.40139 \pm 0.00061 $ & $ 2455969.40138 ^{\mathstrut + 0.00062} _{\mathstrut - 0.00061} $ & $  \phantom{-}0.00047 $ & $ \phantom{-}0.76 $ & Newton-0.20m& clear &2.6	& \texttt{HE6} &ETD obs.~Trnka\\
   129 &2012/02/11& $ 2455969.39959 \pm 0.00054 $ & $ 2455969.39959 ^{\mathstrut + 0.00056} _{\mathstrut - 0.00052} $ & $ -0.00133 $ & $ -2.47 $ & RC-0.25m& clear &2.3	& \texttt{HE7} &ETD obs.~Garcia\\
   145 &2012/03/28& $ 2456015.40859 \pm 0.00034 $ & $ 2456015.40857 ^{\mathstrut + 0.00033} _{\mathstrut - 0.00035} $ & $  \phantom{-}0.00257 $ & $ \phantom{-}7.55 $ & Refractor-0.1m& clear&2.8	 & \texttt{HE8} &ETD obs.~Carre$\tilde{\mathrm{n}}$o\\
\hline
\end{tabular}
}
\note{See Table~\ref{wasp1summary} for a complete description of the columns. The reference ephemeris is according to Eq.\,(\ref{t0_HT1}).}
\end{table*}
%
\section{Ephemeris refinement} \label{ttv}

The construction of an $O-C$ diagram requires a reference ephemeris $T_0 = N\cdot P+ T_\mathrm{ref}$, where $T_0$ is the predicted transit time at epoch $N$ and $T_\mathrm{ref}$ an arbitrary ``zero'' ($N=0$) epoch.

\subsection{WASP-1b}

\begin{figure*}
\centering
\includegraphics[width=0.99\columnwidth, clip]{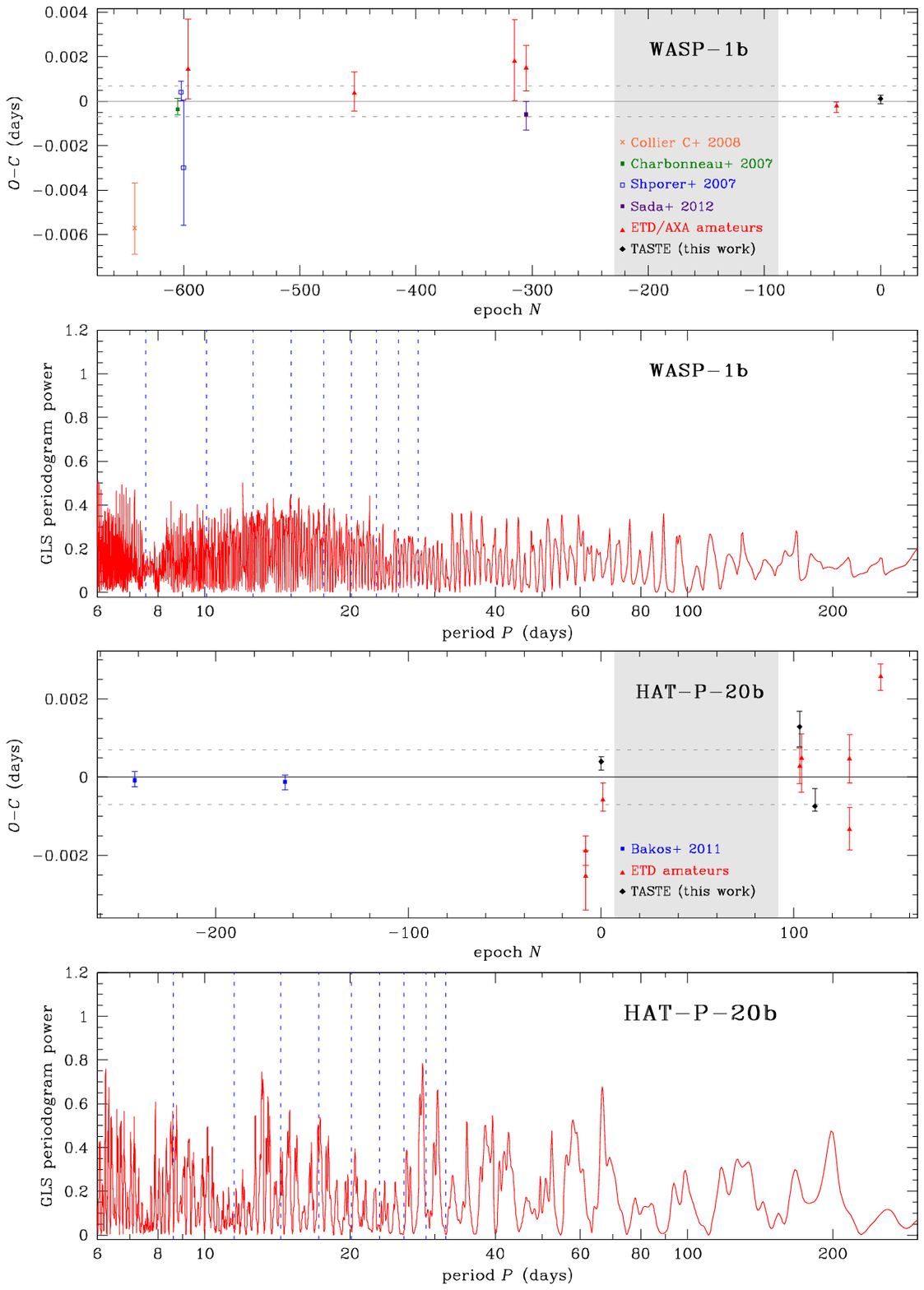}
\includegraphics[width=0.99\columnwidth, clip]{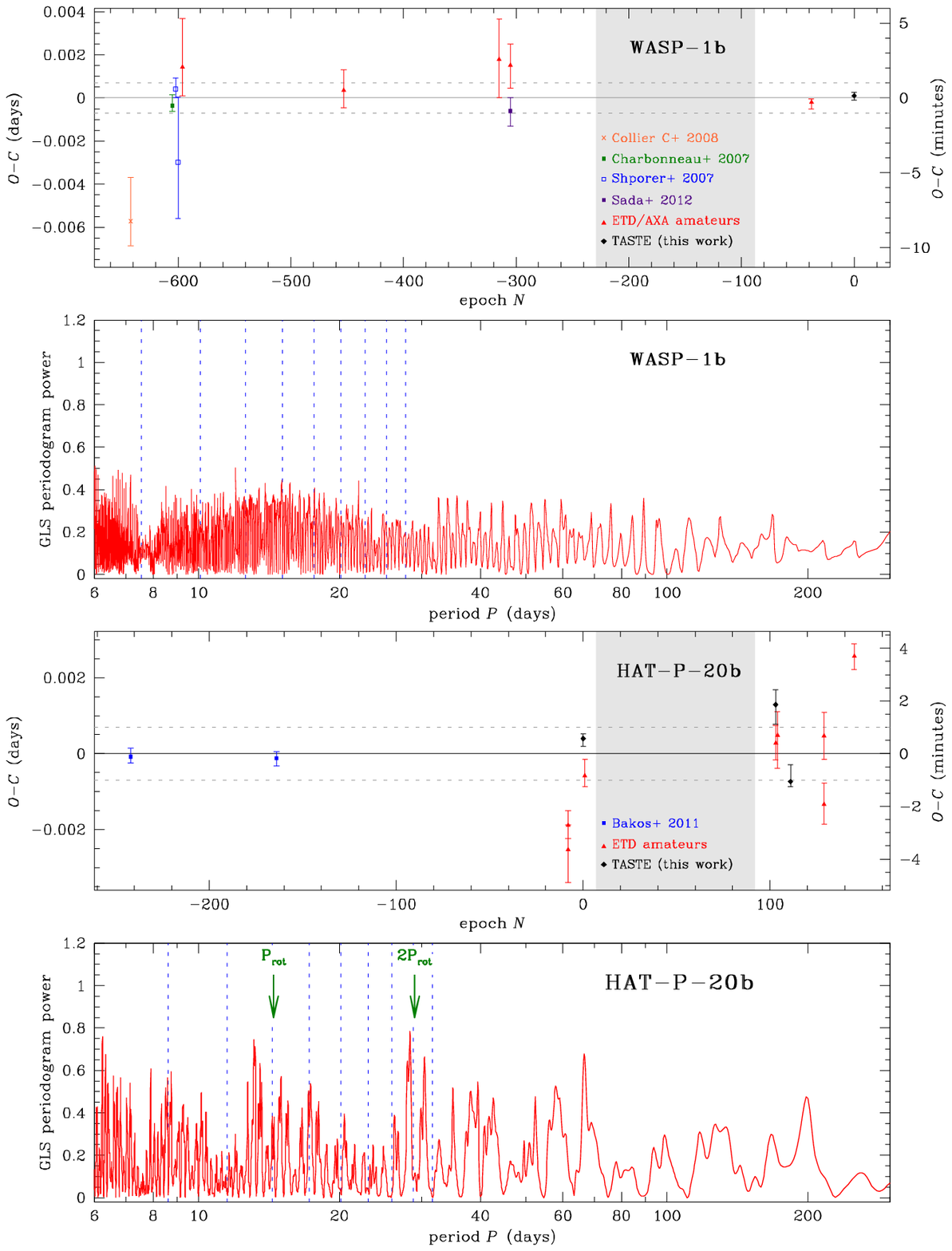}
\caption{\emph{Top-left panel}: $O-C$ diagram for WASP-1b, where every estimated transit time $T_0$ reported in Table~\ref{wasp1summary} has been compared with that predicted by the best-fit ephemeris in Eq.\,(\ref{t0_WT1}) (gray continuous line). Two dashed gray lines mark $\pm 1$ minute from the ephemeris. \emph{Bottom-left panel}: GLS periodogram for all $O-C$ points of WASP-1b, excluding \texttt{WC1}. Blue dashed lines mark the first ten sub-harmonics of the orbital period. \emph{Top- and bottom-right panels}: same as above for HAT-P-20b, employing Eq.\,(\ref{t0_HT1}) as reference ephemeris. The GLS periodogram has been computed from all data points reported in Table~\ref{hat20summary}. The green arrows mark the stellar rotational period $P_\mathrm{rot}$ and its 2:1 sub-harmonic (see detailed discussion in Sec.~\ref{discussion}).}
\label{oc}
\end{figure*}

Employing all the $T_0$ values tabulated in Table~\ref{wasp1summary}, we calculated a new linear ephemeris for WASP-1b by weighted ordinary least squares, setting the new zero point $T_\mathrm{ref}$ at \texttt{WT1}, as this is the most recent and precise transit.
The re-estimated reference ephemeris is:
\begin{equation}
\begin{array}{ccl}  \label{t0_WT1}
T_0\,(\mathrm{BJD}_{\mathrm{TDB}}) & = &(2455530.31970 \pm 0.00016) \,+\\
&& + N\,(2.5199448 \pm 0.0000005)
\end{array}
\end{equation}
where the uncertainties on $P$ and $T_\mathrm{ref}$ were obtained from the covariance matrix of the fit, rescaled by the square root of the reduced $\chi^2$ to take into account the real dispersion of our data. The corresponding $O-C$ data is plotted in the top-left panel of Fig.~\ref{oc}. 
This diagram clearly revealed that the first data point (\texttt{WC1}) is a $3.4\,\sigma$ outlier, probably because of an high level of correlated noise in the light curve. It is worth to mention that, after \texttt{WC1} rejection, $\chi^2_\mathrm{r}\simeq 1.09$ for our fit ($\mathrm{RMS} = 118$~s), indicating that error bars are well estimated and that residuals are in good agreement with a random distribution. The newly inferred $P$ is in agreement with values published in the past, though with a much smaller associated error. 

No TTV signal is detectable for WASP-1b within the noise level allowed by all available data. To confirm this conclusion, we carried out a further analysis by employing the Generalised Lomb-Scargle periodogram (GLS, Zechmeister \& K{\"u}rster 2009). The periodogram plot, for the period range 6--300 days, is shown in the bottom-left panel of Fig.~\ref{oc}; harmonics of the orbital period up to 11:1 are marked with vertical dashed lines. As expected from a white noise distribution, no prominent peak of spectral power is present on the whole frequency range.

\subsection{HAT-P-20b}

The case of HAT-P-20b is more complex. We computed a new reference ephemeris as described above, adopting \texttt{HT1} as $T_\mathrm{ref}$ and employing all the data points reported in Table~\ref{hat20summary}: 
\begin{equation} \label{t0_HT1}
\begin{array}{ccl}
T_0\,(\mathrm{BJD}_{\mathrm{TDB}}) & = & (2455598.48482 \pm 0.00032) \,+\\ 
&& + N\,(2.8753186 \pm 0.000002).
\end{array}
\end{equation}
Again, uncertainties on both quantities were rescaled by $\sqrt{\chi^2_\mathrm{r}}$.  The resulting $O-C$ diagram (Fig.~\ref{oc}, top-right panel) is significantly scattered, being $\chi^2_r \simeq 13$ for the whole data set. Even focusing the analysis only on transits observed by professional astronomers (\texttt{HB1}, \texttt{HB2}, \texttt{HT1}, \texttt{HT2}, and \texttt{HT3}) the observed points are in disagreement with the linear ephemeris model ($\chi^2_r \simeq 11.4$). As discussed in Section~\ref{reduction}, our \texttt{HT2} transit is slightly affected by red noise and this could bias the corresponding value of $T_0$. Nevertheless, both \texttt{HT1} and \texttt{HT3} possess a negligible amount of red noise and both deviate off the  model by $\sim 2.5\,\sigma$, in opposite directions. 
No clear pattern among the residuals is detectable by eye.
The GLS periodogram for the whole HAT-P-20b data set (Fig.~\ref{oc}, bottom-right panel) shows several power peaks, none of them being notably dominant over the other ones.
\begin{table}
\caption{Orbital/physical fitted parameters of WASP-1b and HAT-P-20b estimated from individual light curves. }
\label{param}
\centering
\scalebox{0.70}{
\setlength{\extrarowheight}{8pt}
\begin{tabular}{lccccc}
\multicolumn{6}{c}{WASP-1b} \\
\hline
\raisebox{1ex}[0pt][0pt]{ID}	& \raisebox{1ex}[0pt][0pt]{$(R_\mathrm{p} + R_\star)/a$} & \raisebox{1ex}[0pt][0pt]{$R_\mathrm{p}/R_\star$} & \raisebox{1ex}[0pt][0pt]{$i$} & \raisebox{1ex}[0pt][0pt]{$u_1$} & \raisebox{1ex}[0pt][0pt]{$u_{1,\mathrm{th}}$}\\
\hline
\texttt{WC1} &$0.1978^{\mathstrut +0.0370}_{\mathstrut -0.0124}$ &$0.0985^{\mathstrut +0.0032}_{\mathstrut -0.0030}$ &$88.96^{\mathstrut +0.84}_{\mathstrut -1.39}$ &$1.143^{\mathstrut +0.271}_{\mathstrut -0.417}$ & $0.286$ \\	
\texttt{WH1} &$0.1937^{\mathstrut +0.0066}_{\mathstrut -0.0024}$ &$0.1040^{\mathstrut +0.0009}_{\mathstrut -0.0010}$ &$88.45^{\mathstrut +1.18}_{\mathstrut -1.60}$ &$0.015^{\mathstrut +0.115}_{\mathstrut -0.080}$ & $0.167$ \\
\texttt{WS1} &$0.2009^{\mathstrut +0.0098}_{\mathstrut -0.0069}$ &$0.1049^{\mathstrut +0.0020}_{\mathstrut -0.0021}$ &$86.77^{\mathstrut +1.58}_{\mathstrut -1.50}$ &$0.025^{\mathstrut +0.081}_{\mathstrut -0.112}$ & $0.204$ \\
\texttt{WS2} &$0.1854^{\mathstrut +0.0122}_{\mathstrut -0.0076}$ &$0.1043^{\mathstrut +0.0016}_{\mathstrut -0.0011}$ &$88.40^{\mathstrut +1.25}_{\mathstrut -1.98}$ &$0.122^{\mathstrut +0.097}_{\mathstrut -0.101}$ & $0.204$ \\
\texttt{WD1} &$0.2321^{\mathstrut +0.0243}_{\mathstrut -0.0259}$ &$0.1069^{\mathstrut +0.0014}_{\mathstrut -0.0013}$ &$83.36^{\mathstrut +3.03}_{\mathstrut -2.13}$ &$0.076$	   & $0.076$ \\
\texttt{WT1} &$0.1922^{\mathstrut +0.0016}_{\mathstrut -0.0011}$ &$0.1030^{\mathstrut +0.0007}_{\mathstrut -0.0006}$ &$89.51^{\mathstrut +0.40}_{\mathstrut -0.79}$ &$0.318^{\mathstrut +0.040}_{\mathstrut -0.034}$ & $0.286$ \\
\texttt{WE2} &$0.1953^{\mathstrut +0.0016}_{\mathstrut \mathstrut -0.0015}$ &$0.1029^{\mathstrut +0.0003}_{\mathstrut -0.0003}$ &$89.46^{\mathstrut +0.40}_{\mathstrut -1.03}$ &$0.323$	& $0.323$\\
\hline
\raisebox{1ex}[0pt][0pt]{w.~mean [all $-$ \texttt{WC1}]}  &\raisebox{1ex}[0pt][0pt]{$0.2098^{\mathstrut +0.0024}_{\mathstrut -0.0024}$} &\raisebox{1ex}[0pt][0pt]{$0.1048^{\mathstrut +0.0014}_{\mathstrut -0.0014}$} &\raisebox{1ex}[0pt][0pt]{$86.24^{\mathstrut +2.77}_{\mathstrut -2.77}$} & \raisebox{1ex}[0pt][0pt]{---} & \raisebox{1ex}[0pt][0pt]{---} \\
\hline
\end{tabular}
}
\vskip 4mm
\centering\scalebox{0.71}{
\setlength{\extrarowheight}{8pt}
\begin{tabular}{lccccc}
\multicolumn{6}{c}{HAT-P-20b} \\
\hline
\raisebox{1ex}[0pt][0pt]{ID}	& \raisebox{1ex}[0pt][0pt]{$(R_\mathrm{p} + R_\star)/a$} & \raisebox{1ex}[0pt][0pt]{$R_\mathrm{p}/R_\star$} & \raisebox{1ex}[0pt][0pt]{$i$} & \raisebox{1ex}[0pt][0pt]{$u_1$} & \raisebox{1ex}[0pt][0pt]{$u_{1,\mathrm{th}}$}\\
\hline
\texttt{HB1} &$0.0949^{\mathstrut +0.0055}_{\mathstrut -0.0054}$ &$0.1256^{\mathstrut +0.0020}_{\mathstrut -0.0044}$ &$87.33^{\mathstrut +0.64}_{\mathstrut -0.46}$ &$0.520^{\mathstrut +0.095}_{\mathstrut -0.098}$ & $0.470$ \\
\texttt{HB2} &$0.1023^{\mathstrut +0.0057}_{\mathstrut -0.0054}$ &$0.1287^{\mathstrut +0.0023}_{\mathstrut -0.0032}$ &$86.67^{\mathstrut +0.50}_{\mathstrut -0.52}$ &$0.419^{\mathstrut +0.170}_{\mathstrut -0.178}$ & $0.470$ \\
\texttt{HT1} &$0.1074^{\mathstrut +0.0055}_{\mathstrut -0.0064}$ &$0.1441^{\mathstrut +0.0035}_{\mathstrut -0.0058}$ &$86.46^{\mathstrut +0.62}_{\mathstrut -0.46}$ &$0.663^{\mathstrut +0.131}_{\mathstrut -0.134}$ & $0.595$ \\
\texttt{HT2} &$0.1100^{\mathstrut +0.0080}_{\mathstrut -0.0093}$ &$0.1483^{\mathstrut +0.0027}_{\mathstrut -0.0057}$ &$86.34^{\mathstrut +0.71}_{\mathstrut -0.53}$ &$0.476^{\mathstrut +0.231}_{\mathstrut -0.382}$ & $0.595$ \\
\texttt{HT3} &$0.1105^{\mathstrut +0.0052}_{\mathstrut -0.0051}$ &$0.1456^{\mathstrut +0.0028}_{\mathstrut -0.0041}$ &$86.12^{\mathstrut +0.47}_{\mathstrut -0.49}$ &$0.561^{\mathstrut +0.115}_{\mathstrut -0.215}$ & $0.595$ \\
\texttt{HE1} &$0.1265^{\mathstrut +0.0089}_{\mathstrut -0.0121}$ &$0.1490^{\mathstrut +0.0042}_{\mathstrut -0.0058}$ &$85.29^{\mathstrut +0.80}_{\mathstrut -0.66}$ &$0.595$	& $0.595$ \\
\texttt{HE2} &$0.1079^{\mathstrut +0.0041}_{\mathstrut -0.0048}$ &$0.1377^{\mathstrut +0.0017}_{\mathstrut -0.0020}$ &$86.37^{\mathstrut +0.39}_{\mathstrut -0.27}$ &$0.449$	& $0.449$ \\
\texttt{HE3} &$0.1271^{\mathstrut +0.0082}_{\mathstrut -0.0063}$ &$0.1484^{\mathstrut +0.0037}_{\mathstrut -0.0023}$ &$85.29^{\mathstrut +0.39}_{\mathstrut -0.55}$ &$0.449$	& $0.449$ \\
\texttt{HE4} &$0.1012^{\mathstrut +0.0130}_{\mathstrut -0.0081}$ &$0.1299^{\mathstrut +0.0033}_{\mathstrut -0.0034}$ &$86.83^{\mathstrut +0.70}_{\mathstrut -0.93}$ &$0.449$	& $0.449$ \\
\texttt{HE5} &$0.1155^{\mathstrut +0.0168}_{\mathstrut -0.0132}$ &$0.1244^{\mathstrut +0.0048}_{\mathstrut -0.0046}$ &$85.62^{\mathstrut +1.01}_{\mathstrut -1.05}$ &$0.449$	& $0.449$ \\
\texttt{HE6} &$0.1045^{\mathstrut +0.0101}_{\mathstrut -0.0093}$ &$0.1280^{\mathstrut +0.0034}_{\mathstrut -0.0029}$ &$86.79^{\mathstrut +0.69}_{\mathstrut -0.82}$ &$0.449$	& $0.449$ \\
\texttt{HE7} &$0.1096^{\mathstrut +0.0124}_{\mathstrut -0.0132}$ &$0.1253^{\mathstrut +0.0047}_{\mathstrut -0.0049}$ &$86.16^{\mathstrut +1.03}_{\mathstrut -0.89}$ &$0.449$	& $0.449$ \\
\texttt{HE8} &$0.1324^{\mathstrut +0.0062}_{\mathstrut -0.0080}$ &$0.1450^{\mathstrut +0.0023}_{\mathstrut -0.0024}$ &$85.09^{\mathstrut +0.53}_{\mathstrut -0.42}$ &$0.449$	& $0.449$ \\
\hline
\raisebox{1ex}[0pt][0pt]{w.~mean [all]}  & \raisebox{1ex}[0pt][0pt]{$0.1127^{\mathstrut +0.0108}_{\mathstrut -0.0108}$} & \raisebox{1ex}[0pt][0pt]{---} & \raisebox{1ex}[0pt][0pt]{$86.18^{\mathstrut +0.67}_{\mathstrut -0.67}$} & \raisebox{1ex}[0pt][0pt]{---} & \raisebox{1ex}[0pt][0pt]{---} \\ 
\hline
\end{tabular}
}
\note{The columns give: the ID code of the light curve, the sum of the fractional radii ($R_\mathrm{p} + R_\star)/a$, the ratio $R_\mathrm{p}/R_\star$, the inclination $i$, the linear limb darkening (LD) coefficient $u_1$, and the theoretical linear LD coefficient $u_{1,\mathrm{th}}$ interpolated from the Claret (2000, 2004) tables. The error bars are derived from the \texttt{JKTEBOP} RP algorithm. For amateurs data we fixed $u_1$ coefficient to its theoretical value to avoid non-physical solutions. For both planets, the last row of each sub-table shows the weighted means (see explanation in the text).}
\end{table}
%
\section{Discussion} \label{discussion}

Among the parameters fitted with \texttt{JKTEBOP}, the only inconsistency found with previous literature is on the fractional radius $R_\mathrm{p}/R_\star$ of HAT-P-20b (Table~\ref{param}): its best-fit values for \texttt{HB1} and \texttt{HB2} significantly disagree with those found on \texttt{HT1} and \texttt{HT3}, even if the issue of flux contamination from the neighbor HAT-P-20 B has been taken into account in both works. A smaller inconsistency was found on the linear limb darkening (LD) coefficients: the best-fit values of $u_1$ are slightly smaller in \texttt{HB1} and \texttt{HB2} than those found in \texttt{HT1}, \texttt{HT2} and \texttt{HT3}.

\begin{figure}
\resizebox{\hsize}{!}{\includegraphics[clip]{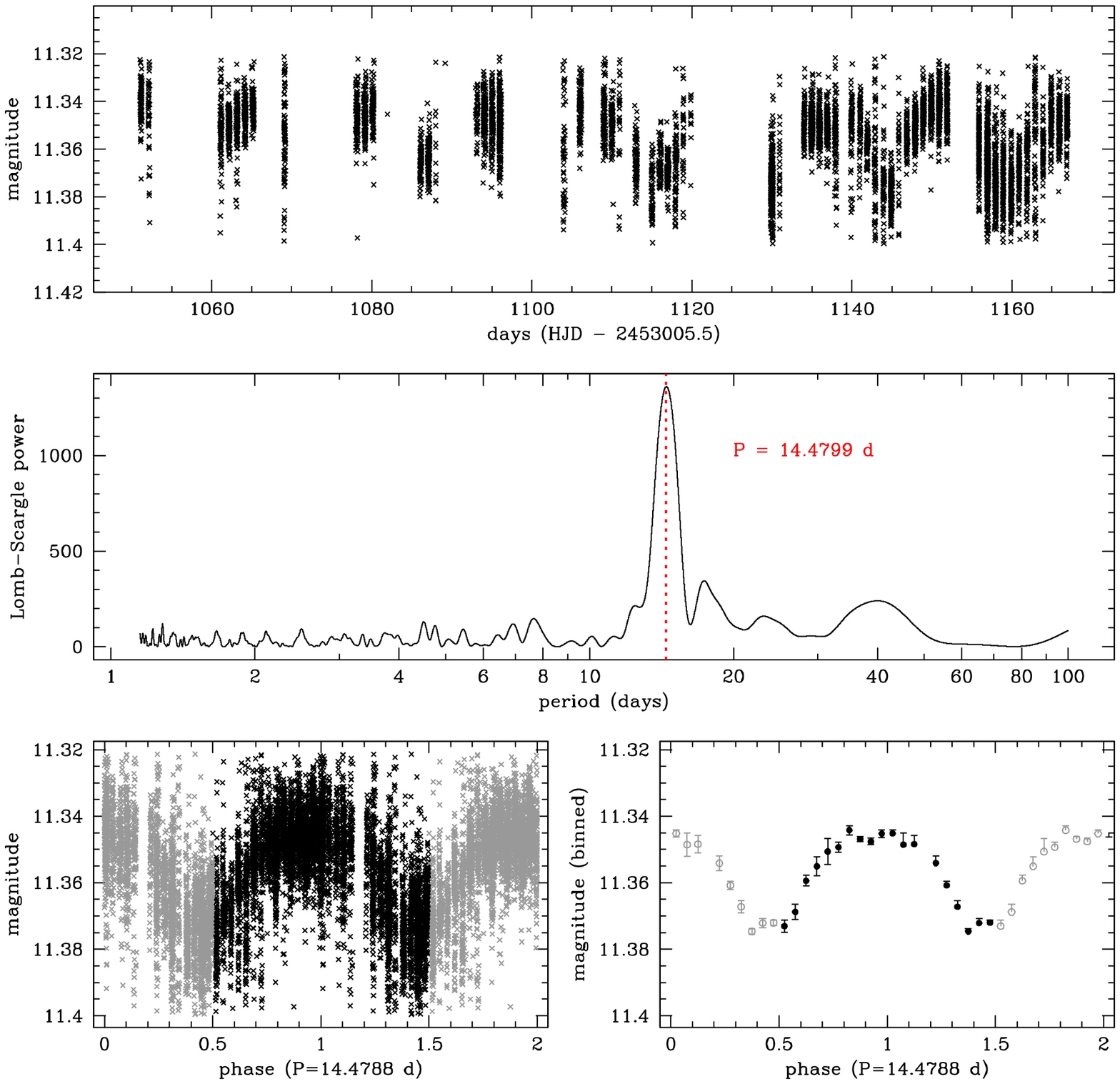}}
\caption{\emph{Top panel}: GLS periodogram of HAT-P-20 from SuperWASP light curve with the marked peak corresponding to a rotational period of $P_{\textrm{rot}}=14.4799$ days. \emph{Bottom panels}: Unbinned (\emph{left}) and binned (\emph{right}) light curve from the same photometric series, folded on the above-mentioned best-fit period. Grey symbols on both panels show repeated phase.}
\label{superwasp}
\end{figure}

Two different scenarios could be invoked to account both for these discrepancies and for the scatter in the $O-C$ plot at epochs $N>0$.
We analyzed the SuperWASP light curve of HAT-P-20~A, downloaded from the public archive\footnote{http://exoplanetarchive.ipac.caltech.edu/docs/\\SuperWASPMission.html} (Pollacco et~al. 2006, Butters et~al. 2010) of the project. The large amount of data points ($>12700$) covers a time span of four years between 2004 and 2008. After applying a GLS periodogram, we detected a single, highly significant peak at $P_{\textrm{rot}} \simeq 14.48$ days (Fig.~\ref{superwasp}, \emph{top panel}).
On such a star, this is a clear indication of photometric modulation caused by stellar rotation and activity; to our knowledge, this is the first of such detection on HAT-P-20~A.
Its relatively short rotational period and the large amplitude of starspots-induced variability ($\sim 0.04$ mag) points toward an age significantly younger than the Sun. Indeed, we compared the $B-V$ color and $P_\mathrm{rot}$ of HAT-P-20~A and the gyrochronological sequence of the $\sim600$~Myr open cluster M44 (Delorme et~al. 2011), obtaining a very good match. As M44 has a metallicity consistent with that of HAT-P-20~A, the age of these two objects might be comparable if one neglects star-planets interactions: if so, the system would look younger and more active than what it is predicted by its real evolutionary state.
At the same time, we examined a high-resolution spectrum of HAT-P-20~A gathered with HIRES at the Keck Observatory (program: N049Hr, PI: Bakos) and noticed extremely strong CaII H\&K emission lines, well above the photospheric continuum, also indicative of a young stellar age. We can conclude that the HAT-P-20 system is probably younger than 1 Gyr, in contrast with the age $6.7^{\mathstrut +5.5}_{\mathstrut -3.8}$ Gyr estimated by Bakos et~al. (2011) through isochrone fitting.
It is worth noting that our detected $P_\textrm{rot}$ lies extremely close to an integer multiple of the orbital period of HAT-P-20b, being $P_\textrm{rot}/P = 5.03$. This is suggestive of a 5:1 resonance driven by the tidal interaction between the spin of the star and its close, massive planetary companion (Szab\'o et~al. 2013). While there is already evidence of other systems locked on similar low-order resonances (Walkowicz \& Basri 2013, Sec 3.3 for a review) the details about the gravitational and magnetic mechanisms involved and the time scale of this phenomenon remain largely unknown (Winn et~al. 2010). Further modeling of the HAT-P-20 system could help in casting light on this field.

From the above-mentioned findings, we suspect that the anomalous scatter in the $O-C$ diagram could be more realistically explained by subtle light curve distortions driven by activity effects on HAT-P-20~A, rather than by a genuine TTV signal or by photometric contamination.
According to this interpretation, \texttt{HB1} and \texttt{HB2} data have been acquired in a relatively low activity phase when a smaller number of spots on the stellar surface makes transits look shallower (Ballerini et~al. 2012, Oshagh et~al. 2013), whereas TASTE and ETD observations occurred at a period of increased stellar activity. The latter period corresponds also to a larger amount of scatter in the $O-C$ diagram, just as expected when starspots are randomly occulted by the planet along the transit path, which lead to a biased estimate of $T_0$ (as in the case of WASP-10b, Barros et~al. 2013).
If this is the case, simultaneous observations of HAT-P-20b in different passbands are required (Ballerini et~al. 2012) to perform an accurate determination of the planetary parameters, and to disentangle the stellar activity contribution from the planetary signal.

An alternative hypothesis to explain the scatter in the $O-C$ plot of HAT-P-20b is the presence of a genuine periodical TTV signal. Folding the $O-C$ diagram on the frequency with the largest spectral power (corresponding to $P$\,$\simeq$\,28.3\,days), the resulting $\chi^2_\mathrm{r}$ value decreases from $\sim$\,13 to $\sim$\,3, suggesting that a pseudo-sinusoidal TTV of semi-amplitude $\sim$\,0.0017\,days ($\sim$\,150\,s) could be in principle a possible explanation for this behaviour.
This peak is at a period about twice as much that found in Fig.~\ref{superwasp} and some power excess is also present at $P_{\textrm{rot}} \simeq 15$; this could further support our speculation about stellar activity. However, we warn that the above-mentioned peak falls on a 10:1 harmonic of the orbital period and that other peaks share a similar level of significance. This is a consequence of the fact that, with only a few high-S/N measurements, our analysis is still based on small-sample statistics.
Any TTV detection is thus only tentative, and our results clearly show that this object deserves further follow-up observations.

\section{Conclusions} \label{conclusion}

In this paper, we presented new TASTE light curves of the hot Jupiters WASP-1b and HAT-P-20b, and performed a homogeneous analysis of both our new data and other archival light curves.
We gave new, more accurate estimate of the physical and orbital parameters of both planets, and carried out an ephemeris refinement to search for variations in the transit periodicity.

While the orbital period of WASP-1b is constant within the estimated uncertainties, the $O-C$ diagram of HAT-P-20b shows an anomalous amount of scatter which could be ascribed to two possible scenarios: 1) a periodical TTV signal or 2) distortions of the light curve due to starspots which affect $T_0$, the fitted transit depth, and the LD coefficients. 
The stellar young age ($< 1$ Gyr) and high level of chromospheric activity of HAT-P-20b, confirmed by both photometry and spectroscopy, and the presence of suspect $O-C$ periodicities close to the rotation period of HAT-P-20A and its integer multiple, lead us to consider the latter hypothesis as more probable.
New simultaneous multi-band photometric observations are needed to confirm or disprove our hypothesis, along with a long-term photometric and spectroscopic follow-up.
\begin{acknowledgements}
We thank the referee Mathias Zechmeister for his careful reading and useful comments and suggestions.
This work was partially supported by PRIN INAF 2008 ``Environmental effects in the formation and evolution of extrasolar planetary system''. V.~G.~acknowledges support from PRIN INAF 2010 ``Planetary system at young ages and the interactions with their active host stars''.  V.~N.~and G.~P.~acknowledge partial support by the Universit\`a di Padova through the ``Progetto di Ateneo \#CPDA103591''. We thank A.~Ayomamitis, A.~Carre$\tilde{\mathrm{n}}$o, R.~Garcia, F.~Hormuth, R.~Naves, J.~Olhert, V.~P$\check{\textrm{r}}$ib\'ik, J.~Trnka (TRESCA) for their light curves extracted from the ETD database and B.~Gary and V.~Hentunen for their data available on NASA Exoplanets Archive. Some tasks of our data analysis have been carried out with the VARTOOLS (Hartman et~al. 2008) and \texttt{Astrometry.net} codes (Lang et~al. 2010).
\end{acknowledgements}

\end{document}